\documentclass[prl,twocolumn,superscriptaddress,showpacs]{revtex4}
\usepackage{graphics}           % for eps figures
\usepackage{bm}                 % for bold math symbols
\usepackage{amsmath}            % for \text and such
\usepackage{verbatim}           % useful for commenting out stuff

\begin{document}
%----------------------------%

%%%%%%%%%%%%%%%%%%% Set up the stuff for the title %%%%%%%%%%%%%%%%%%%%%%%%%%%%
\title{Noiseless Quantum Circuits for the Peres Separability Criterion}
%% {} %% old title

%%% Address labelling is via ordering:

\author{Hilary A. Carteret}
\email{cartereh@iro.umontreal.ca}
\affiliation{Laboratoire d'Informatique Th{\'e}orique et Quantique,
             D{\'e}partement d'Informatique et de Recherche Op{\'e}rationelle,
             Pavillon Andr{\'e}-Aisenstadt, Universit{\'e} de Montr{\'e}al,
             Montr{\'e}al, Qu{\'e}bec, H3C 3J7, Canada}

%%% Reset to actual date before posting %%%
\date{December 8, 2003}

%%% abstract before title in revtex4 %%%
\begin{abstract}
In this Letter we give a method for constructing sets of simple circuits 
that can determine the spectrum of a partially transposed density matrix, 
without requiring either a tomographically complete POVM, or the addition 
of noise to make the spectrum non-negative.  These circuits depend only on 
the dimension of the Hilbert space and are otherwise independent of the state. 
\end{abstract}

%% check PACS again before submitting %%

%----------------------- end of title stuff ----------------------------------%

%%%%%%%%%%%%%%%%%%%%%%%%%%%%%%%%%%%%%%%
%% Actually produce the title first %%%
\maketitle
%-------------------------------------%

%%%%%%%%%%%%%%%%%%%%%%%% Now begin the text %%%%%%%%%%%%%%%%%%%%%%%%%%%%%%%%%%%

\paragraph{Introduction}

There has recently been much interest in measuring functions of the density 
matrix directly, without first performing full state tomography.  The 
non-local properties of density matrices have attracted particular attention
and there has been a series of papers exploring this possibility 
\cite{Filip1,Direct,DirectNLF,DirectLOCC,withoutprior,Filipetal}. 
These typically rely on a combination of the Structural Physical Approximation 
(SPA) \cite{SPA} followed by measuring the spectrum of the resulting density 
operator using the method in \cite{Direct} (but see also 
\cite{Filip1,Filipetal,Toddstuff}).  
The SPA is a method for modifying a map which is not a completely positive 
map (CP map) so that it becomes one.  This is done by forming a convex 
mixture of the original map with another map that projects the state of the 
system onto the maximally mixed state, $M(\rho) = \openone/d,$ for all 
$\rho.$  For maximum sensitivity, we should use as little $M(\rho)$ as 
possible.   For the partial transpose operation on a $d \times d$ dimensional 
system $\openone_A \otimes \Lambda_B,$ this optimal CP-map is given by 
\cite{Direct}
\begin{equation}
 \mathcal{O}= \frac{d^3}{d^3+1}M_A \otimes M_B + 
              \frac{1}{d^3+1}\openone_A \otimes \Lambda_B.
\end{equation}
For a more detailed discussion of the properties of this map and its 
implementation as a positive operator-valued measure, see \cite{FiurasekSPA}.

\begin{figure}[h!]
    \begin{minipage}{\columnwidth}
    \begin{center}
        \resizebox{0.8\columnwidth}{!}{\includegraphics{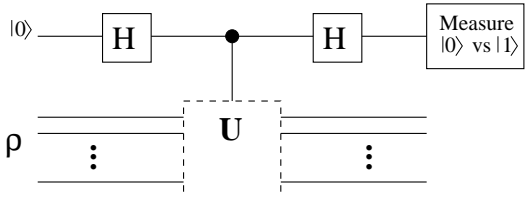}}
    \end{center}
    \end{minipage}
    \caption{General form of an interferometer circuit.}
    \label{general}
\end{figure}

This approach was partly inspired by the interferometer circuits in 
\cite{GPmixed} and the analysis in \cite{KeylW}.  The proof in \cite{GPmixed} 
showed that circuits of the form in figure \ref{general} 
where the density matrices undergo the evolution 
$\bm{\rho} \mapsto U\bm{\rho} U^{\dagger}$ conditioned on the state of the 
control qubit, can measure in the modulation of the interference pattern of 
that qubit, 
\begin{equation}
 \text{Tr}(U{\bm{\rho}})=ve^{i\alpha}
\end{equation}
where $v$ is the visibility, and $\alpha$ is a phase shift.  This is 
significant, because it enables us to characterise the effect of maps 
defined by a unitary map on the kets {\emph{only}}, without being constrained 
to perform $U^{\dagger}$ on the bra vectors. 

If we make $\bm{\rho}=\rho\otimes\rho\otimes\ldots\otimes\rho$ from multiple 
copies of $\rho,$ an $m\times m$ density matrix of interest, we can use this 
circuit to measure the moments of the density matrix, 
Tr$(\rho^2) \ldots$ Tr$(\rho^m)$ by choosing $U$ to be the cyclic shift 
\cite{Direct}
\begin{equation}
 V^{(k)}|\varphi_1\rangle|\varphi_2\rangle\ldots|\varphi_k\rangle =
        |\varphi_k\rangle|\varphi_1\rangle\ldots|\varphi_{k-1}\rangle,
\end{equation}
acting on the {\emph{kets}} of the basis only, thus measuring
\begin{equation}
 \text{Tr}\rho^{\otimes k} V^{(k)} = \text{Tr}\rho^k 
                                   = \sum_{i=1}^m \lambda_i^k,
\end{equation}
from which we can determine the eigenspectrum of the density matrix 
\cite{Filip1,Direct,KeylW}. 

Let us begin by considering polynomial local-unitary-invariants of pure 
states.  These can be constructed as follows.  Begin by writing
\begin{equation}\label{genpoly}
 \psi_{ij\ldots k}\overline{\psi}^{\ell m \dots n} \ldots 
 \psi_{pq\ldots r}\overline{\psi}^{st\ldots u}
\end{equation}
where the tensor indices denote the various subsystems in the usual way.
Now contract every ``downstairs'' (or ``input'') index with one of the 
``upstairs'' (or ``output'') indices labelling the {\emph{same}} subsystem, 
(but not necessarily on a neighbouring term) until no free indices remain. 
These functions are therefore invariant under local basis changes 
{\emph{by construction.}} 

In \cite{Rains97} Rains proved that there was a one-to-one correspondence 
between the set of all local polynomial invariants and sets of permutations 
that permute the downstairs indices corresponding to each subsystem with other 
downstairs indices corresponding to the same subsystem.  (Or equivalently, 
perform the inverse permutation on the upstairs indices.) 
Rains' work was motivated by the need to find new shadow enumerators of 
quantum codes and it was not until \cite{Martin_nCo} that these invariants 
were given a physical interpretation in terms of the expectation values of 
operators.  In \cite{LLW} the methods of \cite{Direct} were extended to 
include arbitrary permutations, thus recovering Rains' construction for 
the full set of polynomial invariants.

The preceding papers {\emph{seem}} to share the assumption that the only way 
to measure the effect of a map on $\rho$ is to cause the density matrix to 
actually undergo that map as a physical evolution: the map must be physically 
implementable.  For maps that are not in this class one must therefore use 
the SPA \cite{SPA}.  (However, we will argue later that a more subtle 
measurement model is implied by certain results in some of these papers.)

This assumption makes intuitive sense and one might think that it is 
therefore ``obviously'' true; except, like some other ``obviously true'' 
assertions about quantum mechanics, it is {\emph{not}}.  To illustrate this, 
we will give a method for constructing a set of unitary circuits that can 
measure the spectrum of $\rho^{{\rm{T}}_2}$ for any $\rho,$ without adding 
noise of any kind.  The structure of these circuits depends only on the 
dimension of $\rho$ and does not require a tomographically complete POVM 
at any stage \cite{FiurasekSPA}.

\paragraph{Measuring the Kempe Invariant}

In \cite{LLW} a method for constructing circuits to measure polynomial 
invariants was presented.  Although the authors do not mention this fact, 
this class of circuits includes a set that can measure the effects of some 
non-positive maps obtained by extending maps that are positive, but not CP 
maps to act on larger spaces. 

\begin{figure}[h!]
    \begin{minipage}{\columnwidth}
    \begin{center}
        \resizebox{0.8\columnwidth}{!}{\includegraphics{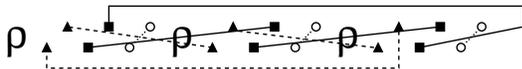}}
    \end{center}
    \end{minipage}
    \caption{The summation diagram for the Kempe invariant.}
    \label{Kempe-rho}
\end{figure}

Consider polynomial invariants of pure states of three qubits.  A $6^{th}$ 
order invariant (in $\psi$) that is algebraically independent of all the 
lower-order invariants was discovered by Kempe \cite{Kempe1} and can be 
written (with a little rearranging) as 
\begin{equation}\label{I6def}
 I_6 = \psi_{k\ell p}\overline{\psi}^{inp}
       \psi_{imq}\overline{\psi}^{j\ell q}
       \psi_{jnr}\overline{\psi}^{kmr}. 
\end{equation}
This index-based notation is rather opaque.  Figure \ref{Kempe-rho} is the 
same invariant drawn as a summation diagram: the linked nodes represent 
summations; each upstairs-downstairs pair of nodes after each copy of 
$\rho$ denotes the indices for one local subsystem (qubit), just as with 
tensor index notation.  The triangles represent indices labelling qubit $1;$ 
the squares, those for qubit $2$ and the open circles correspond to qubit $3.$
\newline
We can always choose to order the copies of $\rho$ in the invariant so that 
one subsystem is traced out; this is just another manifestation of a theorem 
due to Schr{\"{o}}dinger \cite{Schrothm,Mermin,KirkSchro}, which is also known 
as the GHJW theorem \cite{Preskillnotes,GHJW1,GHJW3}.  The traced out party 
is represented by the open circles.  
This way of writing invariants highlights the Rains correspondence between 
invariants and permutations of the indices for each subsystem \cite{Rains97}.  
Consider the following three qubit invariant:
\begin{figure}[h!]
    \begin{minipage}{\columnwidth}
    \begin{center}
        \resizebox{0.8\columnwidth}{!}{\includegraphics{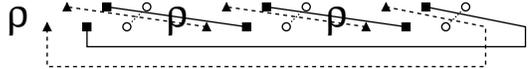}}
    \end{center}
    \end{minipage}
    \caption{Summation diagram for another $3$-qubit invariant.}
    \label{rhocubed}
\end{figure} 
\newline
This is a three qubit invariant but we are interested in bipartite 
entanglement here.   If we ignore the third qubit altogether, we can see that 
this invariant is Tr$((\rho_{12})^3),$ where the subscripts $1,2$ are no 
longer indices, but now refer to qubits $1$ and $2$ and the two longest 
lines represent the final matrix trace.
We are free to drop the third qubit because the indices for particle $3$ are 
always traced out in equation \eqref{I6def} and their links in the summation 
diagram (Fig.~\ref{Kempe-rho}) are never exchanged between copies of $\rho.$

We can now connect figure~\ref{Kempe-rho} to the partial transpose map.  
One way to obtain \ref{Kempe-rho} from \ref{rhocubed} is to take each 
upstairs-downstairs pair of indices for particle $2$ and swap the upstairs 
index with the downstairs one.  This performs the partial transpose operation 
on each copy of $\rho.$  Therefore 
\begin{equation}\label{PT3}
 I_6 = \text{Tr}((\rho_{12}^{{\rm{T}}_2})^3),
\end{equation}
where the symbol $\;^{{\rm{T}}_2}$ denotes the partial transpose with 
respect to qubit $2.$  In other words, equation \eqref{I6def} is the third 
moment of the partial transpose of $\rho.$  

We can measure the Kempe invariant because the summation in 
Fig.~\ref{Kempe-rho} can be obtained from that in Fig.~\ref{rhocubed} in two 
equivalent ways.  The other way is via the Rains construction, using a cyclic 
permutation acting on the downstairs indices for qubit $2$ to reproduce the 
``contraflow'' summation pattern in Fig.~\ref{Kempe-rho}.  As all the indices 
are summed out in upstairs-downstairs pairs, their labels have no physical 
significance and so these two operations produce the same invariant.   Any 
invariant of the form in \eqref{genpoly} with no free indices and all 
summations on upstairs-downstairs pairs can be generated using Rains' 
construction \cite{Rains97}.

\begin{figure}[h!]
    \begin{minipage}{\columnwidth}
    \begin{center}
        \resizebox{0.2\columnwidth}{!}{\includegraphics{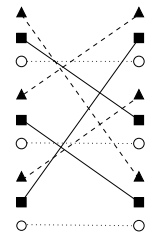}}
    \end{center}
    \end{minipage}
    \caption{The Kempe invariant, in the notation from \cite{LLW}.}
    \label{KempeLLW}
\end{figure}

A circuit can be obtained for this invariant by using the construction in 
\cite{LLW}.  This is done by rotating the summation diagram in figure 
\ref{Kempe-rho} by $90^{\circ}$ clockwise and omitting the $\rho$ symbols  
(see Fig.~\ref{KempeLLW}.)
The downstairs indices now correspond to input rails and the upstairs indices 
to output rails. 
\newline
To construct the circuit, insert the wiring in Fig.~\ref{KempeLLW} as the 
controlled-$U$ in Fig.~\ref{general}.

\paragraph{Measuring the spectrum of $\boldsymbol{\rho^{{\rm{T}}_2}}$}

\begin{figure}[h!]
    \begin{minipage}{\columnwidth}
    \begin{center}
        \resizebox{0.8\columnwidth}{!}{\includegraphics{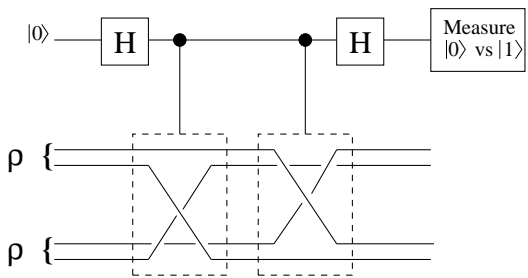}}
    \end{center}
    \end{minipage}
    \caption{The circuit for Tr$((\rho_{12}^{{\rm{T}}_2})^2).$}
    \label{circuit2}
\end{figure}

\begin{figure}[h!]
    \begin{minipage}{\columnwidth}
    \begin{center}
        \resizebox{0.8\columnwidth}{!}{\includegraphics{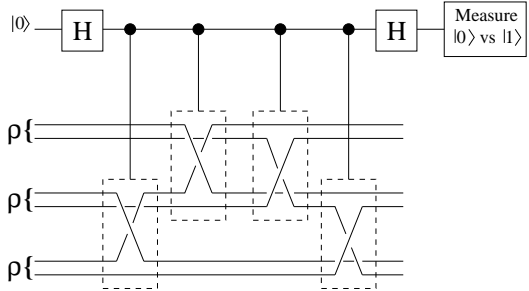}}
    \end{center}
    \end{minipage}
    \caption{The circuit for Tr$((\rho_{12}^{{\rm{T}}_2})^3).$}
    \label{circuit3}
\end{figure}

\begin{figure}[h!]
    \begin{minipage}{\columnwidth}
    \begin{center}
        \resizebox{0.8\columnwidth}{!}{\includegraphics{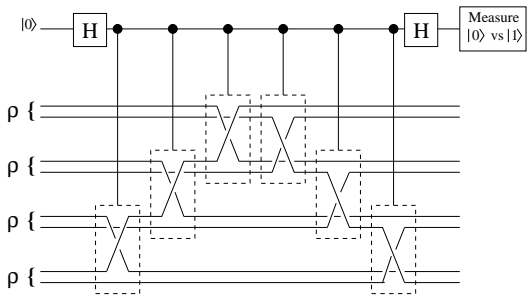}}
    \end{center}
    \end{minipage}
    \caption{The circuit for Tr$((\rho_{12}^{{\rm{T}}_2})^4).$}
    \label{circuit4}
\end{figure}

It can be seen that we can also write the invariants 
Tr$((\rho_{12}^{{\rm{T}}_2})^2)$ and Tr$((\rho_{12}^{{\rm{T}}_2})^4)$ in a 
similar form, so we can construct circuits to evaluate these moments.  We 
can therefore determine the eigenspectrum of $\rho_{12}^{{\rm{T}}_2}$ using 
the recipe given in \cite{Direct,KeylW} (or alternatively, reconstructing the 
characteristic polynomial for $\rho_{12}^{{\rm{T}}_2}$ from the values of 
Tr$((\rho_{12}^{{\rm{T}}_2})^k)).$  Thus we do not need to use the SPA 
\cite{SPA} to determine the spectrum of $\rho^{{\rm{T}}_2}.$  The circuits 
for the second, third and fourth moments are shown in 
Figs.~\ref{circuit2}-\ref{circuit4} respectively, with the permutations 
decomposed into controlled-{\small{SWAP}} ({\small{CSWAP}}) gates.   
These are the only circuits required for two qubit states; we do not need a 
circuit for Tr$(\rho_{12}^{{\rm{T}}_2})$ because the partial transpose is a 
trace preserving map, so the first moment is already known to be $1.$  
In fact, the only unitary circuit we can write down that is first order 
in $\rho,$ or indeed $\rho_{12}^{{\rm{T}}_2}$ is the circuit for the trace 
norm.  This is also the only polynomial invariant that is quadratic in $\psi.$

\paragraph{Discussion}

We have exhibited a class of circuits that can measure the spectrum of 
$\rho^{{\rm{T}}_2},$ without needing to perform a tomographically complete
POVM or add noise.  While we have only given the circuits for the two 
qubit case, the construction generalizes to bipartite systems of any finite 
dimension in the obvious way; all that is required is two cyclic permutations 
of opposite handedness acting on the rails for each subsystem.  
Fiur{\'{a}}{\u{s}}ek \cite{FiurasekSPA} has shown that the SPA for the 
partial transpose for two qubits is a tomographically complete CP map.  It 
is an open problem whether this is generally true for the partial transpose 
on larger systems, but the method given in this paper never needs a 
tomographically complete POVM to characterize the effects of the partial 
transpose on $\rho,$ regardless of its dimension.  These circuits are 
surprisingly simple: if we can implement an eigenspectrum measurement by 
the method in \cite{Filip1,Direct} then we can measure the spectrum of 
$\rho^{{\rm{T}}_2}$ with a set of circuits that uses the same set of 
primitives.  The complexity of the networks also scales the same way 
as those for eigenspectrum circuits; for a general $d_1 \times d_2$ 
dimensional mixed bipartite state, the circuits consume the same number of 
copies per run, $\tfrac{1}{2}d_1^2d_2^2+\tfrac{1}{2}d_1d_2-1.$

What may be more interesting about these circuits is that they disprove our 
intuition that {\emph{the only characterizable map is a physically 
implementable map.}}  This statement is still true if we only have access to 
one copy of the state at a time, as this method for measuring the effects of 
maps needs multiple copies in order to work; since we cannot implement the 
map directly, we must characterise it via its moments.  It is also worth 
noting that having access to only two copies at a time is not much use 
either; the circuit for Tr$((\rho_{12}^{{\rm{T}}_2})^2)$ is identical to 
that for Tr$(\rho_{12}^2).$  Thus the partial transpose may also preserve 
Tr$(\rho^2).$ 

If we can manipulate at least $d_1d_2$ copies at once, we can perform the 
Peres test \cite{Perestest}.  This is a necessary and sufficient condition 
for separability for two qubit states, although it is not sufficient to test 
for separability in general \cite{H3sep}.  The entire minimal generating set 
for the ring of polynomial invariants described by Rains \cite{Rains97,LLW} 
is necessary to completely separate the orbits of the state under local 
unitary operations in general.  It is also sufficient 
\cite{VnO,QIProb3,3qubits}.  However, there is no known upper bound on the 
order of the invariants needed to construct a complete minimal generating set.
One way to see this is to notice that this paper has only examined invariants 
whose summation pattern has period $1$ in the number of copies of the 
density matrix it passes through before it repeats.  There is no 
{\emph{a priori}} reason why there might not be invariants corresponding to 
longer-period summation patterns that cannot be written as products of terms 
with period $1$ and thus could have a much larger order.  

This is not the only reason why this problem is hard, but it is enough to 
show that a general density operator may need circuits in more than 
$d_1d_2$ copies in order to characterize its non-local properties completely.  
We cannot say how many copies we would need to be able to process at once 
without first finding the minimal generating set of invariants for the system. 
This is a hard problem and all we can say in general is that this number is 
finite \cite{Springer}.

If we assume that we can use unitary circuits that can manipulate any finite 
number of copies at once, the class of measurable maps is considerably larger 
than the set of all unitary maps.  Consider a density matrix that we have the 
option of pre-processing according to 
\begin{equation}\label{preprocess}
 \rho^{\otimes n} \mapsto V\rho^{\otimes n} V^{\dagger}
\end{equation}
for any unitary $V.$  If our interferometer circuit contains a controlled-$U$ 
of our choice, we can now measure any 
\begin{equation}\label{WUV}
 \rho^{\otimes n} \mapsto W \rho^{\otimes n} V^{\dagger}
\end{equation}
where $W=UV,$  which is a much larger class of operations than is accessible 
by equation \eqref{preprocess} alone.  What is not immediately obvious is that 
this class of operations is qualitatively different from the usual unitary 
operations.  This is because we now have the ability to operate on the ket 
basis of the density matrix {\emph{without}} having to perform the adjoint 
operation on the bra basis. 
\newline

%%%%%%%%%%%%%%%%%%%%%%%%%%%%%%%%%%%%%%%%%%%%%%%%%%%%%%%%%%%%%%%%%%%%%%%%%%%%%%

%---------------------%
\begin{acknowledgments}
%---------------------%

I would like to thank Martin R{\"{o}}tteler, Vlatko Vedral, Todd Brun and 
the IQIS group at the University of Calgary for interesting discussions.  
This research was supported by MITACS, The Fields Institute and the 
NSERC CRO project ``Quantum Information and Algorithms.''

%---------------------%
\end{acknowledgments}
%---------------------%

%%%%%%%%%%%%%%%%%%%%%%%%%%%%%%%%%%%%%%%%%%%%%%%%%%%%%%%%%%%%%%%%%%%%%%%%%%%

%-------------------------------------%
%%%% construct refs with natbib... %%%%
%\bibliography{gphaseptx}
%----------------%

%-------------------------------------------%
%%% ...then replace with .bbl file at end %%%
%---------------------------%

%----------------------%
%%% must end with... %%%
\end{document}